\documentclass{ws-ijmpcs}
\usepackage[usenames]{color}
\def\trimmarks{}

\newcommand{\dhd}{{\textstyle d}
\lower.03ex\hbox{\kern-0.40em$^{\scriptstyle-}$}\kern-0.08em{}}  

\newcommand{\dbar}{{\textstyle \delta}
\lower.03ex\hbox{\kern-0.38em$^{\scriptstyle-}$}\kern-0.05em{}}

\newcommand{\bu}{{\bullet}}

\newcommand{\cald}{{\cal D}}  
\newcommand{\calf}{{\cal F}}  
  
\newcommand{\calo}{{\cal O}}

\newcommand{\tilcaf}{\tilde{\cal F}} 
\newcommand{\ticalo}{\tilde{\cal O}}

\newcommand{\tilA}{\tilde{A}} 
 
\newcommand{\tilT}{\tilde{T}} 
\newcommand{\tilU}{\tilde{U}}

\begin{document}

\markboth{I. Balitsky, A. Tarasov}
{Evolution  of gluon  TMD}

%
\catchline{}{}{}{}{}
%

\title{Evolution of gluon TMD at low and moderate $x$}

\author{I. BALITSKY and A.TARASOV}

\address{Physics Dept., ODU, Norfolk VA 23529\\
and \\
Theory Group, Jlab, 12000 Jefferson Ave, Newport News, VA 23606,USA\\
E-mail: balitsky@jlab.org, atarasov@jlab.org}

\maketitle


\begin{abstract}
We study how the rapidity evolution of gluon transverse momentum dependent distribution 
changes from nonlinear evolution at small $x\ll 1$ to linear double-logarithmic evolution at moderate $x\sim 1$.
\end{abstract}

\ccode{PACS numbers: 12.38.Bx,  12.38.Cy}

\section{Introduction\label{aba:sec1}}

\bigskip
A TMD factorization [\refcite{jimayuan}, \refcite{collinsbook}] generalizes the usual concept of parton density by allowing PDFs to
depend on intrinsic transverse momenta in addition to the usual longitudinal momentum fraction variable. 
These transverse-momentum dependent parton distributions (also called unintegrated parton distributions) are widely used in the analysis of 
semi-inclusive processes like Drell-Yan process or single-inclusive deep inelastic scattering (SIDIS) 
or dijet production in hadron-hadron collisions (for a review, see Ref. [\refcite{collinsbook}]). However, the analysis of TMD evolution in these cases are mostly
restricted to the evolution of quark TMDs, whereas at high collider energies the majority of produced particles
will be small-$x$ gluons. In this case one has to understand the transition between non-linear dynamics at small $x$ and
presumably linear evolution of gluon TMDs at intermediate $x$, which is not clear at present.

To make things even more complicated, there are  two non-equivalent definitions of gluon TMDs in small-$x$ and ``medium $x$'' communities.
In the small-$x$ literature the Weiszacker-Williams (WW) unintegrated gluon distribution [\refcite{domarxian}] is defined in terms of the matrix element 
\begin{equation}
\sum_X {\rm tr}\langle p| U\partial^iU^\dagger(z_\perp)|X\rangle \langle X| U\partial_iU^\dagger(0_\perp)\} |p\rangle
\end{equation}
between target states (typically protons). Here $\sum_X$ denotes the sum over full set of hadronic states and $U_z$ is a Wilson-line operator - infinite gauge link ordered along the light-like line
\begin{equation}
U(z_\perp)~=~[\infty n+z_\perp,-\infty n+z_\perp],~~~~[x,y]~\equiv~Pe^{ig\int\! du~(x-y)^\mu~A^a_\mu(ux+(1-u)y)}
\label{defu}
\end{equation}
In the spirit of rapidity factorization, Bjorken $x$ enters this expression as a rapidity cutoff for Wilson-line operators. 
Roughly speaking, each gluon emitted by Wilson line has rapidity restricted from above by $\ln x_B$.
 
One can rewrite the above matrix element (up to some trivial factor) in the form
\begin{eqnarray}
&&\hspace{-0mm}
\alpha_s\cald(x_B,z_\perp)~=~-{\alpha_s\over 2\pi(p\cdot n)x_B}\!\int\! du \sum_X \langle p|\tilcaf^a_\xi(z_\perp+un)|X\rangle \langle X| \calf^{a\xi}(0)|p\rangle
\nonumber\\
&&\hspace{-0mm}
\calf^a_\xi(z_\perp+un)~\equiv~[\infty n+z_\perp,u n+z_\perp]^{am}n^\mu F^m_{\mu \xi}(un+z_\perp)
\nonumber\\
&&\hspace{-0mm}
\tilcaf^a_\xi(z_\perp+un)~\equiv~n^\mu F^m_{\mu \xi}(un+z_\perp)[un+z_\perp,\infty n+z_\perp]^{ma}
\label{dgTMD}
\end{eqnarray}
and define the ``WW unintegrated gluon distribution''
\begin{equation}
\cald(x_B,k_\perp)~=~\!\int\!d^2z_\perp~e^{-i(k,z)_\perp}\cald(x_B,z_\perp)
\end{equation}
where $(k,z)_\perp$ denotes the scalar product in 2-dim transverse Euclidean space.
It should be noted that since Wilson lines are renorm-invariant $\alpha_s\cald(x_B,k_\perp)$ does not depend on the renormalization scale $\mu$.

On the other hand, at moderate $x_B$ the unintegrated gluon distribution is defined as [\refcite{muldrod}]
\begin{eqnarray}
&&\hspace{-0mm}
\cald(x_B,k_\perp,\eta)~=~\!\int\!d^2z_\perp~e^{-i(k,z)_\perp}\cald(x_B,z_\perp,\eta),
\label{gTMD}\\
&&\hspace{-0mm}
\alpha_s\cald(x_B,z_\perp,\eta)~=~{-x_B^{-1}\alpha_s\over 2\pi(p\cdot n)}\!\int\! du ~e^{-ix_Bu(pn)}\sum_X \langle p|\tilcaf^a_\xi(z_\perp+un)|X\rangle \langle X| \calf^{a\xi}(0)|p\rangle
\nonumber
\end{eqnarray}
There are more involved definitions with Eq. (\ref{gTMD}) multiplied by some Wilson-line factors [\refcite{collinsbook}] but we will discuss the ``primordial'' TMD (\ref{gTMD}).
The Bjorken $x$ is now introduced explicitly in the definition of gluon TMD. However, because light-like Wilson lines exhibit rapidity divergencies, we need a 
separate cutoff $\eta$ (not necessarily equal to $\ln x_B$) for the rapidity of the gluons emitted by Wilson lines. In addition,
 the matrix elements (\ref{gTMD}) will have double-logarithmic contributions 
of the type $(\alpha_s\eta\ln x_B)^n$ while the WW distribution (\ref{dgTMD}) has only single-log terms $(\alpha_s \ln x_B)^n$ 
described by the BK evolution [\refcite {npb96,yura}].

 In the present paper we study the connection between rapidity evolution of two gluon TMDs  (\ref{dgTMD}) and (\ref{gTMD})
 in the kinematic region where $s\gg k_\perp^2$ and $k_\perp^2\geq $ few GeV$^2$. The latter condition means that we can use perturbative QCD while the former one indicates that 
 TMD factorization, rather than the collinear factorization, is applicable [\refcite{collinsbook}]. In this kinematic region we will vary Bjorken $x$ and look how non-linear evolution at small $x$ transforms
 into linear evolution at moderate $x_B$.
 
It should be emphasized that we consider gluon TMD with Wilson links going to $+\infty$ in the longitudinal direction.  
Another gluon TMD with links going to $-\infty$ arises in the problems with exclusive particle production, see the discussion in Ref. [\refcite{muxiyu}].
 We plan to study it in future publications.
 
 The paper is organized as follows. In Sec. 2 we remind the general logic of rapidity factorization and rapidity evolution. 
 In Sec. 3 we calculate the Lipatov vertex of the gluon production by TMD operator and in Sec.  4 we obtain the so-called
 virtual corrections. The obtained TMD evolution is discussed in Sec. 5.

\section{Rapidity factorization and evolution \label{sec2}}

In the spirit of high-energy OPE, the rapidity of the gluons is restricted from above by the ``rapidity divide'' $\eta$ separating the impact factor and the matrix element so the proper definition of $U_x$ is 
\begin{eqnarray}
&&\hspace{-0mm} 
 U^\eta_x~=~{\rm Pexp}\Big[ig\!\int_{-\infty}^\infty\!\! du ~p_1^\mu A^\eta_\mu(up_1+x_\perp)\Big],
\nonumber\\
&&\hspace{-0mm} 
A^\eta_\mu(x)~=~\int\!{d^4 k\over 16\pi^4} ~\theta(e^\eta-|\alpha|)e^{ik\cdot x} A_\mu(k)
\label{cutoff}
\end{eqnarray}
where  the  Sudakov variable $\alpha$ is defined as usual,  $k=\alpha p_1+\beta p_2+k_\perp$.
We define the light-like vectors $p_1$ and $p_2$ such that $p_1=n$  and $p_2=p-{m^2\over s}n$ where $p$ is the momentum of the target particle of mass $m$. 
We use metric $g^{\mu\nu}~=~(1,-1,-1,-1)$ so $p\cdot q~=~(\alpha_p\beta_q+\alpha_q\beta_p){s\over 2}-(p,q)_\perp$. For the coordinates we use 
the notations $x_\bu\equiv x_\mu p_1^\mu$ and $x_\ast\equiv x_\mu p_2^\mu$ related to the light-cone coordinates by $x_\ast=\sqrt{s\over 2}x_+$ and $x_\bu=\sqrt{s\over 2}x_-$.
It is convenient to define Fourier transform of the operator $\calf^a_i$
\begin{eqnarray}
&&\hspace{-0mm} 
\calf^a_i(k_\perp,\beta_B)~=~\!\int\! d^2z_\perp~e^{-i(k,z)_\perp}\calf^a_i(z_\perp,\beta_B),
\nonumber\\
&&\hspace{-0mm} 
\calf^a_i(z_\perp,\beta_B)~\equiv~{2\over s}
\!\int\! dz_\ast ~e^{i\beta_B z_\ast} 
[\infty,z_\ast]_z^{am}F^m_{\bu i}(z_\ast,z_\perp)
\label{kalf}
\end{eqnarray}
and similarly
\begin{eqnarray}
&&\hspace{-0mm} 
\tilcaf^a_i(k_\perp,\beta_B)~=~\!\int\! d^2z_\perp~e^{i(k,z)_\perp}\tilcaf^a_i(z_\perp,\beta_B),
\nonumber\\
&&\hspace{-0mm} 
\tilcaf^a_i(z_\perp,\beta_B)~\equiv~{2\over s}
\!\int\! dz_\ast ~e^{-i\beta_B z_\ast} 
F^m_{\bu i}(z_\ast,z_\perp)[z_\ast,\infty]_z^{ma}
\label{tilkaf}
\end{eqnarray}
in the complex-conjugate part of the amplitude. 
Here we introduced the ``Bjorken $\beta_B$''  to decouple this notation from $x_B$ in the denominator in the r.h.s. of Eq. (\ref{gTMD}). 
Also, hereafter we use the notation $[\infty, z_\ast]_z\equiv[\infty p_1+z_\perp, {2\over s}z_\ast p_1+z_\perp]$  where  
$[x,y]$ stands for the straight-line gauge link connecting points $x$ and $y$ as defined in Eq. (\ref{defu}).

In this notations the unintegrated gluon TMD (\ref{gTMD}) can be represented as 
\begin{eqnarray}
&&\hspace{-0mm} 
\langle p|\tilcaf^a_i(k'_\perp,\beta'_B)\calf^{ai}(k_\perp,\beta_B)|p\rangle 
\equiv\sum_X\langle p|\tilcaf^a_i(k'_\perp,\beta'_B)|X\rangle\langle X|\calf^{ai}(k_\perp,\beta_B)|p\rangle 
\nonumber\\
&&\hspace{-0mm} 
=~-2\pi\delta(\beta_B-\beta'_B) (2\pi)^2\delta^{(2)}(k_\perp-k'_\perp)
2\pi x_B\cald(\beta_B=x_B,k_\perp,\eta)
\label{TMD}
\end{eqnarray}
Hereafter we use a short-hand notation
\begin{eqnarray}
&&\hspace{-0mm} 
\langle p|\ticalo_1...\ticalo_m\calo_1...\calo_n |p\rangle
\equiv~\sum_X\langle p| \tilT\{ \ticalo_1...\ticalo_m\}|X\rangle\langle X|T\{\calo_1...\calo_n\}|p\rangle
\label{ourop}
\end{eqnarray}
where tilde on the operators in the l.h.s. of this formula stands as a reminder that they should be inverse time ordered
as indicated by inverse-time ordering $\tilT$ in the r.h.s. of the above equation.

As discussed e.g. in Refs [\refcite{keld}] such martix element can be represented by a double functional integral
\begin{eqnarray}
&&\hspace{-0mm} 
\langle \ticalo_1...\ticalo_m\calo_1...\calo_n \rangle
\nonumber\\
&&\hspace{-0mm} 
=~\int\! D\tilA D\tilde{\bar\psi}D\tilde{\psi}~e^{-iS_{\rm QCD}(\tilA,\tilde{\psi})}\!\int\! DA D\bar{\psi} D\psi ~e^{iS_{\rm QCD}(A,\psi)} \ticalo_1...\ticalo_m\calo_1...\calo_n
\label{funtegral}
\end{eqnarray}
with the boundary condition $\tilA(\vec{x},t=\infty)=A(\vec{x},t=\infty)$ (and similarly for quark fields) reflecting the sum over all intermediate states $X$. 
Due to this condition, the matrix element (\ref{TMD}) can be made gauge-invariant by connecting the endpoints of Wilson lines at infinity with the gauge link
\begin{eqnarray}
&&\hspace{-0mm} 
\langle p|\tilcaf^a_i(z'_\perp,\beta'_B)\calf^{ai}(z_\perp,\beta_B)|p\rangle~
\nonumber\\
&&\hspace{-0mm} 
\rightarrow~\langle p|\tilcaf^a_i(z'_\perp,\beta'_B)[z'_\perp+\infty p_1,z_\perp+\infty p_1]\calf^{ai}(z_\perp,\beta_B)|p\rangle
\label{inftylink}
\end{eqnarray}
This gauge link is important if we use the light-like gauge $p_1^\mu A_\mu=0$ for calculations [\refcite{bejuan}], but in all other gauges it can be neglected.
We will not write it down explicitly but will always assume it in our formulas.

In the spirit of rapidity factorization, in order to find the evolution of the operator $\tilcaf^a_i(z'_\perp,\beta'_B)\calf^{ai}(z_\perp,\beta_B)$ with respect to rapidity cutoff $\eta$, 
see Eq. (\ref{cutoff}), one should integrate in the matrix element  (\ref{TMD}) over gluons and quarks with rapidities $\eta>Y>\eta'$ and temporarily ``freeze'' fields with $Y<\eta'$ to be integrated over later. (For a review, see Refs. [\refcite{mobzor,nlolecture})]. In this case, we obtain functional integral of Eq. (\ref{funtegral}) type over fields with $\eta>Y>\eta'$ in the ``external'' fields with $Y<\eta'$. 
In terms of Sudakov variables we integrate over gluons with $\alpha$ between $\sigma_1=e^\eta$ and $\sigma_2=e^{\eta'}$ and, in the leading order, only
the diagrams with gluon emissions are relevant - the quark diagrams will enter as loops at the next-to-leading (NLO) level.

\begin{figure}[htb]
\begin{center}
\includegraphics[width=34mm]{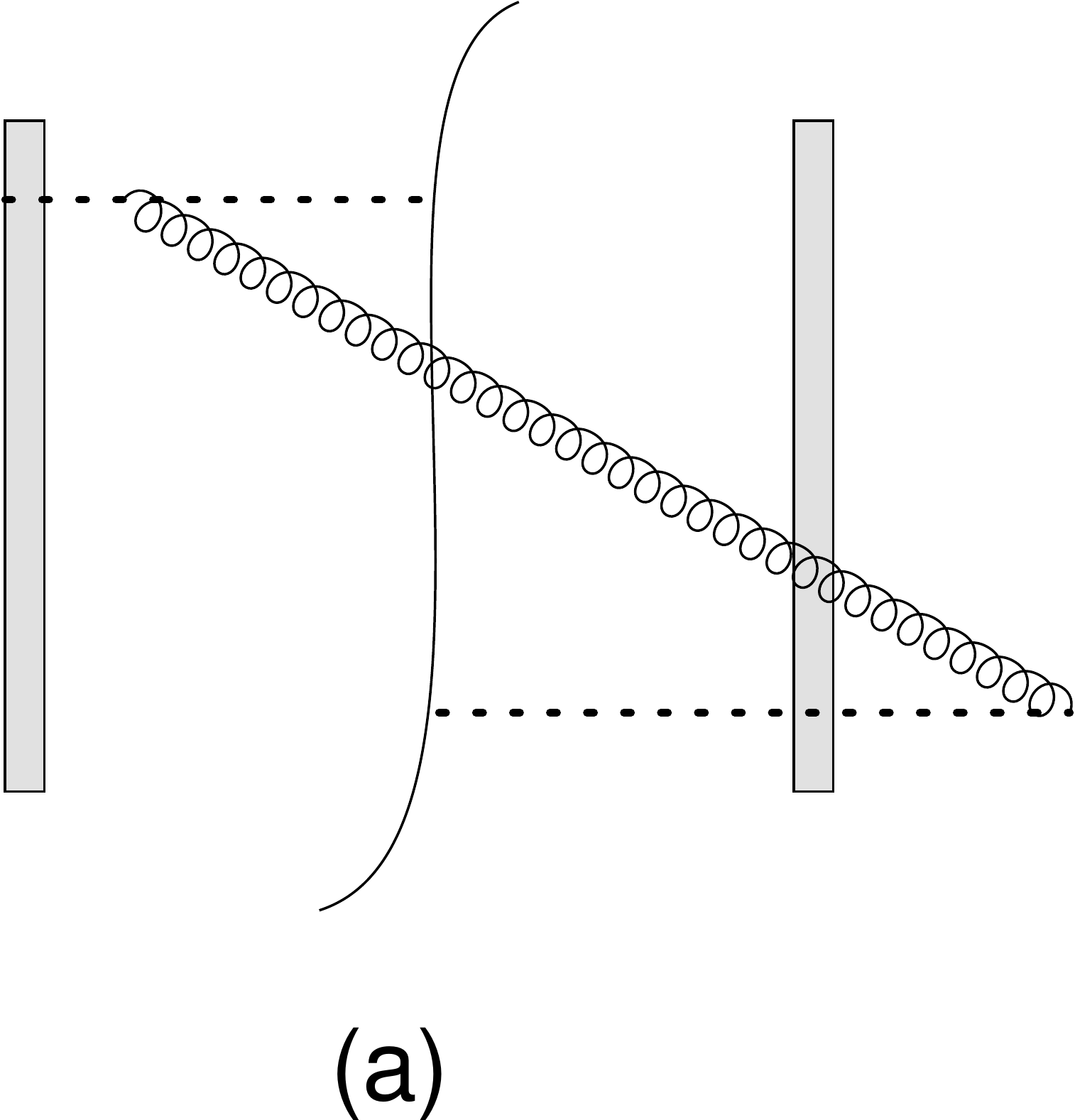}
\hspace{22mm}
\includegraphics[width=34mm]{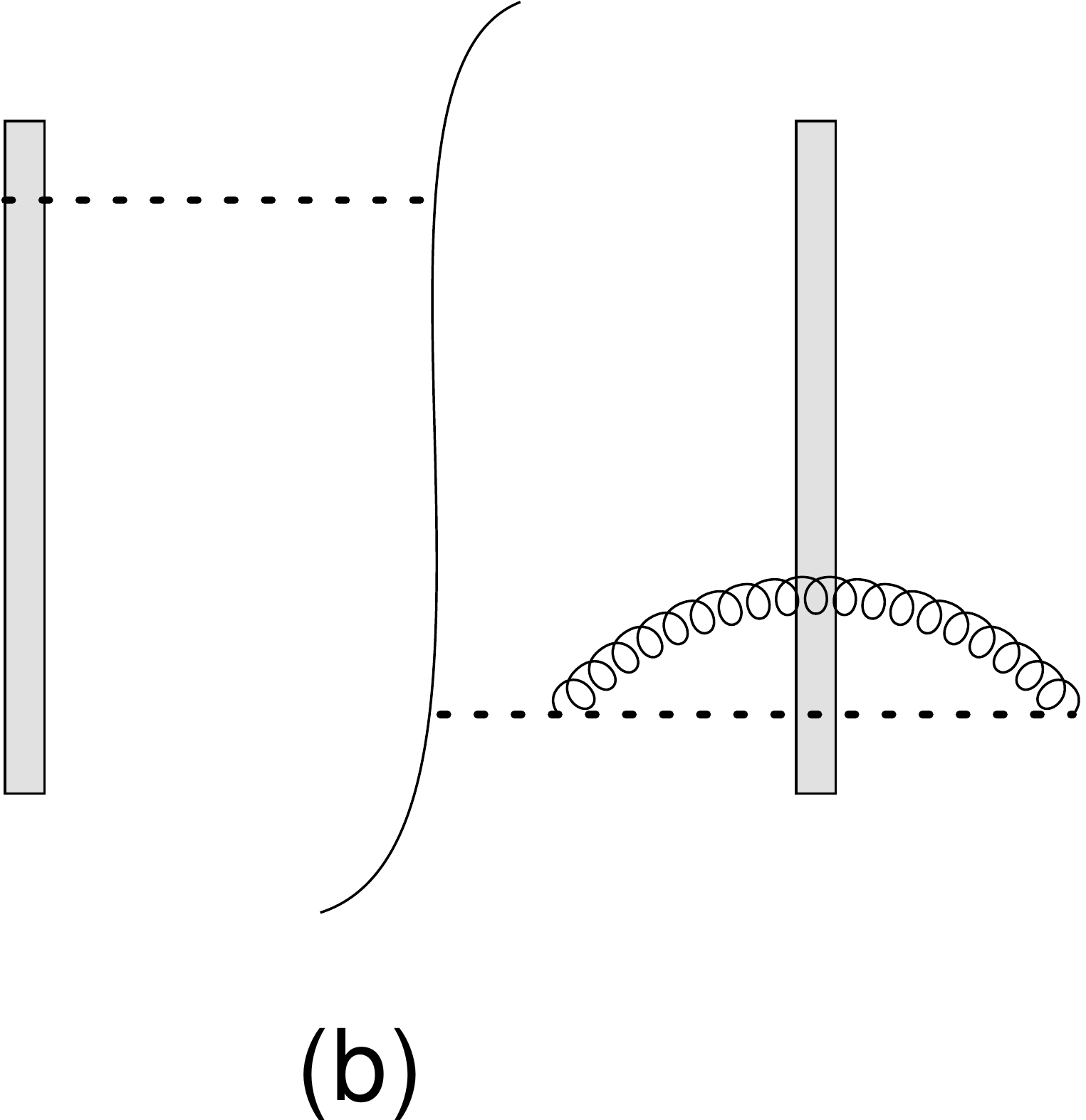}
\end{center}
\caption{Typical diagrams for real (a) and virtual (b) contributions to the evolution kernel. The dashed lines denote gauge links.\label{fig:1}}
\end{figure}
As discussed in Ref. [\refcite{npb96}],  the interaction of gluons with large $\alpha$ with small $\alpha$ fields is described by eikonal  gauge factors. The typical longitudinal size of small $\alpha$ 
fields is $\sigma_\ast\sim {\sigma_2 s\over m^2}$, which is much smaller than the typical distances $\sim{\sigma_1 s\over m^2}$ traveled by large-$\alpha$ gluons. Effectively,
large-$\alpha$ gluons propagate in the external field of the small-$\alpha$ shock wave. In the leading order there is only one extra gluon and we get the typical diagrams of Fig. 1 type. 
The real part of the kernel can be obtained as a square of a Lipatov vertex - the amplitude of the emission of a real gluon by the operator $\calf^a_i$.

\subsection{Lipatov vertex}

 In this Section we will present the real contribution
corresponding to square of Lipatov vertex describing the emission of a gluon from the operator $\calf^a_i$.
The Lipatov vertex is defined as
\begin{eqnarray}
&&\hspace{-11mm}
L^{ab}_\mu(k,z_\perp)~=~i\lim_{k^2\rightarrow 0}k^2\langle T\{A^a_\mu(k)\calf^b_i(z_\perp,\beta_B)\}\rangle
\label{fla9}
\end{eqnarray}
It is convenient to use light-like gauge $p_2^\mu A_\mu=0$, cf. Ref [\refcite{nlobk}]. 
The three corresponding diagrams are shown in Fig. 2. 
\begin{figure}[htb]
\begin{center}
\includegraphics[width=88mm]{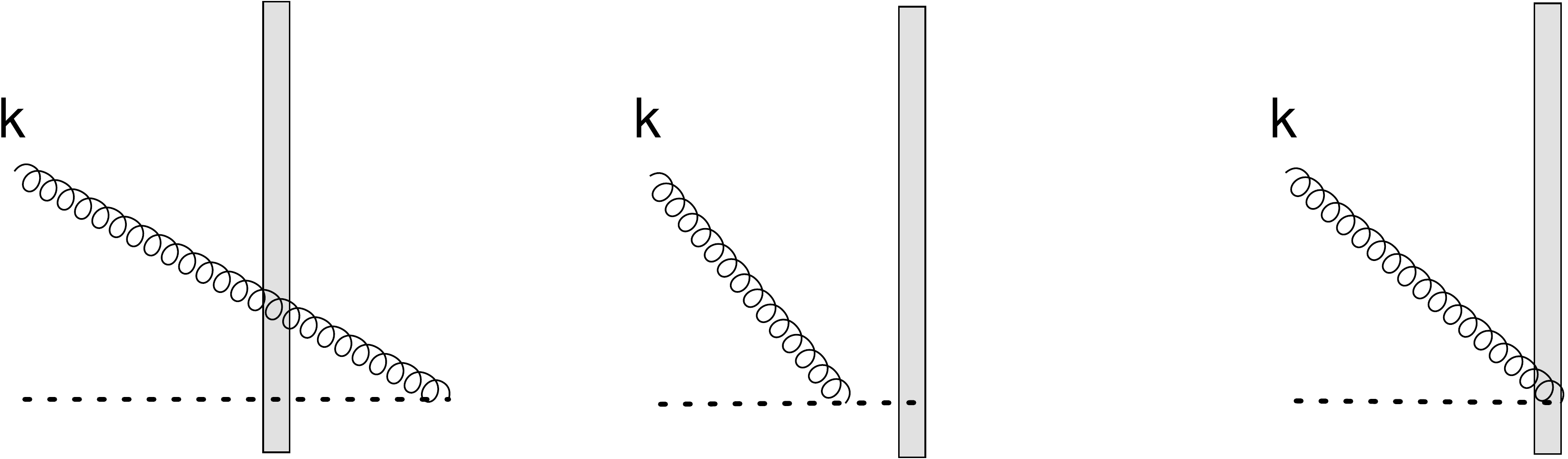}
\end{center}
\caption{Lipatov vertex of gluon emission. \label{fig:2}}
\end{figure}

The actual calculations will be published elsewhere and here we present only the final results in the  $p_2^\mu A_\mu=0$ gauge
\begin{eqnarray}
&&\hspace{-11mm}
L^{ab}_\mu(k,z_\perp)~=~i\lim_{k^2\rightarrow 0}k^2\langle T\{A^a_\mu(k)\calf^b_i(z_\perp,\beta_B)\}\rangle
\nonumber\\
&&\hspace{-11mm}
=~\alpha\beta_Bs\big(g^{\perp}_{\mu i}-{2\over s\alpha}p_{2\mu}k^{\perp}_i\big)
(k_\perp|{1\over p_\perp^2+\alpha\beta_Bs}-U{1\over p_\perp^2+\alpha\beta_Bs}U^\dagger |z_\perp)^{ab}
\nonumber\\
&&\hspace{-11mm}
+~2(k_{\perp}|U{p_i\big[p^\perp_\mu+{2\over\alpha s}p_{2\mu}(p,k)_\perp\big]\over p_\perp^2+\alpha\beta_Bs}U^\dagger
-{p_i\big[p^\perp_\mu+{2\over\alpha s}p_{2\mu}(p,k)_\perp\big]\over p_\perp^2+\alpha\beta_Bs}|z_{\perp})^{ab}
\nonumber\\
&&\hspace{51mm}
+~2(k_\perp|\big({p^\perp_\mu\over p_\perp^2}+{2p_{2\mu}\over\alpha s}\big) \calf_i(\beta_B)|z_\perp)^{ab}
\label{lvresult}
\end{eqnarray}
where we use  Schwinger's notations
$$
(x_\perp|f(p_\perp)|y_\perp)~\equiv~ \int\! \dhd^2p_\perp~e^{i(p,x-y)_\perp}f(p), ~~~~~(x_\perp|p_\perp)~=~e^{i(p,x)_\perp}
$$ 
and the space-saving notation $\dhd^np\equiv {d^np\over(2\pi)^n}$.
It is easy to check that $k^\mu L^{ab}_\mu(k,z_\perp)~=~0$ as required by gauge invariance.

The cross section of gluon emission will be proportional to the square of Lipatov vertex (\ref{lvresult}). 
The terms $\sim p_{2\mu}$ do not contribute to the square so one  can use the Lipatov vertex of transverse gluon emission in the form
\begin{equation}
\hspace{-0mm}
L^{ab}_{\mu_\perp}(k,q_\perp)~
=~(k_{\perp}|U{p_\perp^2g^\perp_{\mu i}+2p_ip^\perp_\mu\over p_\perp^2+\alpha\beta_B s}U^\dagger
-{p_\perp^2g^\perp_{\mu i}+2p_ip^\perp_\mu\over p_\perp^2+\alpha\beta_B s}
+2{p^\perp_\mu\over p_\perp^2} \calf_i(\beta_B)|z_\perp)^{ab}
\label{lvertaxial}
\end{equation}
It is worth noting that at $\beta_B=0$ this results agrees with the Lipatov vertex obtained in Ref. [\refcite{bal04}]
The real-production part of the evolution kernel for operator (\ref{TMD}) is the square of  the Lipatov vertex (\ref{lvertaxial}):
\begin{eqnarray}
&&\hspace{-0mm} 
\langle \tilcaf^{ai}(z'_\perp,\beta_B)\calf^a_i(z_\perp,\beta_B)\rangle_{\rm real}
\nonumber\\
&&\hspace{-0mm} 
=~-{g^2\over 4\pi}\!\int_{\sigma_2}^{\sigma_1}\!{d\alpha\over\alpha}{\rm Tr}
(z'_\perp|\Big[\tilU{p_\perp^2g_{ij}+2p_ip_j\over p_\perp^2+\alpha\beta_B s}\tilU^\dagger
-{p_\perp^2g_{ij}+2p_ip_j\over p_\perp^2+\alpha\beta_B s}
+2\tilcaf_i(\beta_B){p_j\over p_\perp^2} \Big]
\nonumber\\
&&\hspace{-1mm}
\times~
\Big[U{p_\perp^2g^{ij}+2p^ip^j\over p_\perp^2+\alpha\beta_B s}U^\dagger
-{p_\perp^2g^{ij}+2p^ip^j\over p_\perp^2+\alpha\beta_B s}
+2{p^j\over p_\perp^2} \calf^i(\beta_B)\Big]|z_\perp)
\label{realresult}
\end{eqnarray}
where Tr stands for the trace in the adjoint representation and the summation goes over the transverse Latin indices $i,j=1,2$. 
This expression is UV finite and the IR-divergent contribution has the form
\begin{equation}
{g^2\over \pi}N_c\!\int_{\sigma_2}^{\sigma_1}\!{d\alpha\over\alpha}\tilcaf^{ai}(\beta_B,z'_\perp)
(z'_\perp|{1\over p_\perp^2}|z_\perp)\calf^a_i(\beta_B,z_\perp)
\label{IReal}
\end{equation}

\section{Virtual corrections}
\begin{figure}[htb]
\begin{center}
\includegraphics[width=99mm]{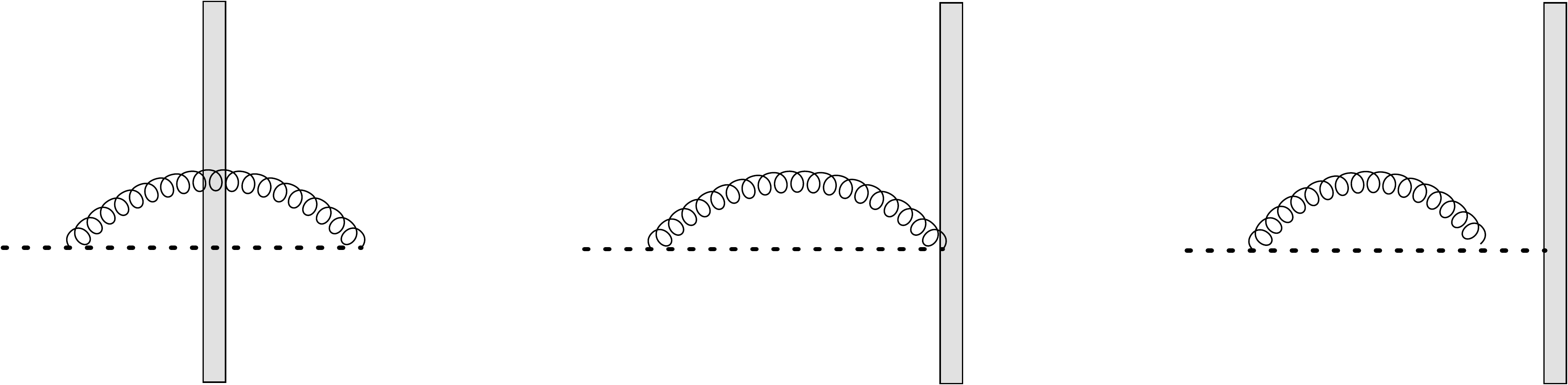}
\end{center}
\caption{Virtual gluon corrections. \label{fig:3}}
\end{figure}

Apart from the ``real correction'' part of the evolution kernel proportional to square of Lipatov vertex (\ref{realresult}) we should take into account  virtual corrections coming from diagrams 
shown in Fig. \ref{fig:3}.
The result for these diagrams is
\begin{eqnarray}
&&\hspace{-1mm} 
\langle \calf^a_i(\beta_B,z_\perp)\rangle ~=~{g^2\over 2\pi}\!\int_{\sigma_2}^{\sigma_1}\!{d\alpha\over\alpha}
\Big\{-N_c\calf^a_i(\beta_B,z_\perp)~(z_\perp|{1\over p_\perp^2}|z_\perp)
\label{virtual}\\
&&\hspace{-1mm}
+~if^{abc}\Big[(z_\perp|{1\over p_\perp^2}\calf_i(\beta_B)U{p_\perp^2\over\alpha\beta_B s+p_\perp^2}U^\dagger|z_\perp)^{bc}
-(z_\perp|{1\over p_\perp^2}\partial_\perp^2U{p_i\over\alpha\beta_B s+p_\perp^2}U^\dagger|z_\perp)^{bc}\Big]\Big\}
\nonumber\\
&&\hspace{-1mm}
=~-i{g^2\over 2\pi}f^{abc}\!\int_{\sigma_2}^{\sigma_1}\!{d\alpha\over\alpha}(z_\perp|
{1\over p_\perp^2}\calf_i(\beta_B)U{\alpha\beta_Bs\over\alpha\beta_Bs+p_\perp^2}U^\dagger
+
{1\over p_\perp^2}\partial_\perp^2U{p_i\over\alpha\beta_B s+p_\perp^2}U^\dagger
|z_\perp)^{bc}
\nonumber
\end{eqnarray}
It should be emphasized that the UV divergence over $p_\perp^2$ in the r.h.s.  of this equation is absent due to the cutoff  $\alpha<\sigma_1$. 

The total virtual correction is the sum of  the correction (\ref{virtual}) with  the similar correction in the complex conjugate amplitude
\begin{eqnarray}
&&\hspace{-1mm} 
\langle \tilcaf^{ai}(\beta_B,z'_\perp)\calf^a_i(\beta_B,z_\perp)\rangle 
~=~-{g^2\over 2\pi}\!\int_{\sigma_2}^{\sigma_1}\!{d\alpha\over\alpha}
\label{virtual1}\\
&&\hspace{-1mm}
\times~{\rm Tr}\big\{\tilcaf^i(\beta_B,z'_\perp)
(z_\perp|{1\over p_\perp^2}\calf_i(\beta_B)U{\alpha\beta_Bs\over\alpha\beta_Bs+p_\perp^2}U^\dagger
+{1\over p_\perp^2}\partial_\perp^2U{p_i\over\alpha\beta_B s+p_\perp^2}U^\dagger|z_\perp)
\nonumber\\
&&\hspace{-1mm}
+~(z'_\perp|\tilU{p_i\over\alpha\beta_B s+p_\perp^2}
\partial_\perp^2\tilU^\dagger{1\over p_\perp^2}
+\tilU{\alpha\beta_Bs\over\alpha\beta_Bs+p_\perp^2}\tilU^\dagger\tilcaf_i(\beta_B){1\over p_\perp^2}|z'_\perp)\calf^i(\beta_B,z_\perp)
\big\}
\nonumber
\end{eqnarray}
Note that he IR divergence in the r.h.s. of this equation cancels with that of Eq. (\ref{IReal}).

\section{Discussion}
Adding together the production part (\ref{realresult}) and the virtual corrections (\ref{virtual}) one obtains
\begin{eqnarray}
&&\hspace{-1mm} 
\langle \tilcaf^{ai}(z'_\perp,\beta_B)\calf^a_i(z_\perp,\beta_B)\rangle
\label{result}\\
&&\hspace{-1mm} 
=~{g^2\over 4\pi}{\rm Tr}\!\int_{\sigma_2}^{\sigma_1}\!{d\alpha\over\alpha}\Big\{(z'_\perp|
-\Big[\tilU{p_\perp^2g^\perp_{ij}+2p_ip_j\over p_\perp^2+\alpha\beta_B s}\tilU^\dagger
-{p_\perp^2g_{ij}+2p_ip_j\over p_\perp^2+\alpha\beta_B s}
+2\tilcaf_i(\beta_B){p_j\over p_\perp^2} \Big]
\nonumber\\
&&\hspace{21mm}
\times~
\Big[U{p_\perp^2g^{ij}+2p^ip^j\over p_\perp^2+\alpha\beta_B s}U^\dagger
-{p_\perp^2g^{ij}+2p^ip^j\over p_\perp^2+\alpha\beta_B s}
+2{p^j\over p_\perp^2} \calf^i(\beta_B)\Big]|z_\perp)
\nonumber\\
&&\hspace{-1mm}
-~2\tilcaf^i(z'_\perp,\beta_B)\calf_i(z_\perp,\beta_B)\Big[(z_\perp|{1\over p_\perp^2}|z_\perp)+(z'_\perp|{1\over p_\perp^2}|z'_\perp)\Big]
+2\tilcaf^i(z'_\perp,\beta_B)
\nonumber\\
&&\hspace{-1mm}
\times~(z_\perp|{p^j\over p_\perp^2}U{p_\perp^2 g_{ij}+2p_ip_j\over\alpha\beta_B s+p_\perp^2}U^\dagger|z_\perp)
+2(z'_\perp|\tilU{p_\perp^2 g_{ij}+2p_ip_j\over\alpha\beta_B s+p_\perp^2}\tilU^\dagger{p^j\over p_\perp^2}|z'_\perp)\calf^i(z_\perp,\beta_B)\Big]
\Big\}
\nonumber
\end{eqnarray}
It can be rewritten in the form where the cancellation of UV and IR divergencies is obvious
\begin{eqnarray}
&&\hspace{-1mm} 
\langle \tilcaf^{ai}(z'_\perp,\beta_B)\calf^a_i(z_\perp,\beta_B)\rangle
\label{result1}\\
&&\hspace{-1mm} 
=~{g^2\over 4\pi}{\rm Tr}\!\int_{\sigma_2}^{\sigma_1}\!{d\alpha\over\alpha}\Big\{
-(z'_\perp|\Big[\tilU{p_\perp^2g_{ij}+2p_ip_j\over p_\perp^2+\alpha\beta_B s}\tilU^\dagger
-{p_\perp^2g_{ij}+2p_ip_j\over p_\perp^2+\alpha\beta_B s}\Big]
\nonumber\\
&&\hspace{41mm}
\times~\Big[U{p_\perp^2g^{ij}+2p^ip^j\over p_\perp^2+\alpha\beta_B s}U^\dagger
-{p_\perp^2g_{ij}+2p_ip_j\over p_\perp^2+\alpha\beta_B s}\Big]|z_\perp)
\nonumber\\
&&\hspace{-1mm}
+~2(z'_\perp|\big[\tilU{p^i\over p_\perp^2+\alpha\beta_B s}\tilU^\dagger-{p^i\over p_\perp^2+\alpha\beta_B s}\big]\calf_i(\beta_B)
\nonumber\\
&&\hspace{41mm}
+~\tilcaf^{i}(\beta_B)\big[U{p_i\over p_\perp^2+\alpha\beta_B s}U^\dagger-{p_i\over p_\perp^2+\alpha\beta_B s}\big]|z_\perp)
\nonumber\\
&&\hspace{-1mm}
+~2\tilcaf^{i}(z'_\perp,\beta_B)\big[
(z'_\perp|{1\over p_\perp^2}\partial_\perp^2U{p_i\over\alpha\beta_B s+p_\perp^2}U^\dagger|z_\perp)
-(z_\perp|{1\over p_\perp^2}\partial_\perp^2U{p_i\over\alpha\beta_B s+p_\perp^2}U^\dagger|z_\perp)\big]
\nonumber\\
&&\hspace{-1mm}
+~2\big[(z'_\perp|\tilU{p_i\over\alpha\beta_B s+p_\perp^2}\partial_\perp^2\tilU^\dagger{1\over p_\perp^2}|z_\perp)
-(z'_\perp|\tilU{p_i\over\alpha\beta_B s+p_\perp^2}\partial_\perp^2\tilU^\dagger{1\over p_\perp^2}|z'_\perp)\big]
\calf^i(z_\perp,\beta_B)
\nonumber\\
&&\hspace{-1mm}
+~2\tilcaf^{i}(\beta_B,z'_\perp)\big[(z'_\perp|{1\over p_\perp^2}\calf_i(\beta_B)
U{\alpha\beta_Bs\over\alpha\beta_B s+p_\perp^2}U^\dagger|z_\perp)
\nonumber\\
&&\hspace{41mm}
-~(z_\perp|{1\over p_\perp^2}\calf_i(\beta_B)U{\alpha\beta_Bs\over\alpha\beta_B s+p_\perp^2}U^\dagger|z_\perp)\big]
\nonumber\\
&&\hspace{-1mm}
+~2\big[(z'_\perp|\tilU{\alpha\beta_Bs\over \alpha\beta_Bs+p_\perp^2}\tilU^\dagger\tilcaf^i(\beta_B){1\over p_\perp^2}|z_\perp)
\nonumber\\
&&\hspace{41mm}
-~
(z'_\perp|\tilU{\alpha\beta_Bs\over \alpha\beta_Bs+p_\perp^2}\tilU^\dagger
\tilcaf^{i}(\beta_B){1\over p_\perp^2}|z'_\perp)\big]\calf_i(\beta_B,z_\perp)\Big\}
\nonumber
\end{eqnarray}
Note that in the transition from Eq. (\ref{result}) to Eq. (\ref{result1}) we dropped contributions $\sim{m^2\over\alpha\beta_Bs}\big(1-e^{i\beta_Bz_\ast}\big)$ for $z_\ast$ lying inside the shock wave since they are small at any $\beta_B$.

It is worth noting that the same formula (\ref{result1}) will hold true for the fragmentation TMD defined as 
\begin{eqnarray}
&&\hspace{-0mm} 
\langle \tilcaf^a_i(k'_\perp,\beta'_B)\calf^{ai}(k_\perp,\beta_B)\rangle_{\rm frag}
\equiv\sum_X\langle 0|\tilcaf^a_i(k'_\perp,\beta'_B)|p+X\rangle\langle p+X|\calf^{ai}(k_\perp,\beta_B)|0\rangle 
\nonumber\\
&&\hspace{-0mm} 
=~-2\pi\delta(\beta_B-\beta'_B) (2\pi)^2\delta^{(2)}(k_\perp-k'_\perp)
2\pi x_B\cald_{\rm frag}(\beta_B=x_B,k_\perp,\eta)
\label{TMD}
\end{eqnarray}

It is easy to see that our formula (\ref{result1}) for the evolution kernel smoothly interpolates between the $k_T$-factorization and TMD-factorization cases.
Indeed, in the framework of the usual small-$x$ approximation $\beta_B$ is neglected so the corresponding ``small-x'' gluon TMD looks like
\begin{equation}
\hspace{-0mm}
\calf^a_i(z_\perp,0)~=~U^a_i(z_\perp)~\equiv~-2i{\rm tr}\{t^aU\partial_iU^\dagger\}
\end{equation}
(where tr stands for the trace in the fundamental representation) 
and Eq. (\ref{result}) reduces to the non-linear equation
\begin{eqnarray}
&&\hspace{-1mm} 
\langle \tilU^{ai}(z'_\perp) U^a_i(z_\perp)\rangle
\label{result1}\\
&&\hspace{-1mm} 
=~{g^2\over \pi}\!\int_{\sigma_2}^{\sigma_1}\!{d\alpha\over\alpha}\Big[-
(z'_\perp|\tilU p_i\tilU^\dagger\big(\tilU{p_j\over p_\perp^2}\tilU^\dagger 
-{p_j\over p_\perp^2}\big)\big(U{p^j\over p_\perp^2}U^\dagger 
-{p^j\over p_\perp^2}\big)Up^i U^\dagger|z_\perp)^{aa}
\nonumber\\
&&\hspace{-1mm}
-~{N_c\over 2}\tilU^a_i(z'_\perp)U^{ai}(z_\perp)\big[(z_\perp|{1\over p_\perp^2}|z_\perp)+(z'_\perp|{1\over p_\perp^2}|z'_\perp)\big]
\nonumber\\
&&\hspace{-1mm}
+~if^{abc}\tilU^a_i(z'_\perp)(z_\perp|{p^j\over p_\perp^2}U{p^ip_j\over p_\perp^2}U^\dagger|z_\perp)^{bc}
+if^{abc}(z'_\perp|\tilU{p^ip_j\over p_\perp^2}\tilU^\dagger{p^j\over p_\perp^2}|z'_\perp)^{ab}U^c_i(z_\perp)\Big]
\nonumber
\end{eqnarray}
which agrees with the results of Ref. [\refcite{mobzor}]. 
It is convenient to rewrite this equation as follows (cf. Ref. [\refcite{domumuxi}]):
\begin{eqnarray}
&&\hspace{-1mm} 
{d\over d\eta}\tilU^a_i(z_2) U^a_i(z_1)
\label{result2}\\
&&\hspace{-1mm}
=~-{g^2\over 8\pi^3}{\rm Tr}
(-i\partial^{z_2}_i+\tilU^{z_2}_i)\big[\!\int\! d^2z_3(\tilU_{z_2}\tilU^\dagger_{z_3}-1)
{z_{12}^2\over z_{13}^2z_{23}^2}(U_{z_3}U^\dagger_{z_1}-1)\big]
(i\stackrel{\leftarrow}{\partial^{z_1}_i}+U^{z_1}_i)
\nonumber 
\end{eqnarray}
where all indices are 2-dimensional and Tr stands for the trace in the adjoint representation. 
It is easy to see that the expression in the square brackets is actually the BK kernel for the double-functional integral 
for cross sections [\refcite{mobzor,difope}].

On the other hand, if $\beta_B\sim 1$ so that $\alpha\beta_Bs\gg p_\perp^2$ we get a linear equation
\begin{eqnarray}
&&\hspace{-1mm} 
\langle \tilcaf^{ai}(z'_\perp,\beta_B)\calf^a_i(z_\perp,\beta_B)\rangle
\label{result3}\\
&&\hspace{-1mm} 
=~-{g^2N_c\over \pi}\!\int_{\sigma_2}^{\sigma_1}\!{d\alpha\over\alpha}
\!\int\!{\dhd^2 p\over p^2}\big[1-e^{i(p,z'-z)_\perp}\big]
\langle\tilcaf^{ai}(z'_\perp,\beta_B)\calf^a_i(z_\perp,\beta_B)\rangle
\nonumber
\end{eqnarray}
which can be rewritten as a linear equation
\begin{eqnarray}
&&\hspace{-1mm} 
{d\over d\eta}\cald(x_B,z_\perp,\eta)
~=~-{\alpha_sN_c\over \pi^2}\cald(x_B,z_\perp,\eta)
\!\int\!{d^2 p\over p^2}\big[1-e^{i(p,z)_\perp}\big]
\nonumber
\end{eqnarray}

We see that the IR divergence at $p_\perp^2\rightarrow 0$ cancels while the UV divergence in the virtual correction
should be cut from above by the condition $p_\perp^2< \sigma s$ following from Eq. (\ref{result}). Actually, at  $x_B\sim 1$
there will be logarithmical region $e^\eta m\sqrt{s} \gg p_\perp^2\gg m^2$ so one has to sum up leading logarithms  $\big(\alpha_s\eta\big)^n$ 
in the evolution kernel Eq. (\ref{result3}) after which the kernel should reproduce the usual Sudakov double logarithms. 

From Eq. (\ref{result}) it is clear that the transition between linear evolution (\ref{result3}) and the non-linear evolution 
(\ref{result2}) occurs at $x_B=\beta_B\sim {m^2\over s}$. We plan to study this transition in future publications.

The authors are grateful to G.A. Chirilli, J.C. Collins,  A. Prokudin,  A.V. Radyushkin, and F. Yuan for valuable discussions. This work was supported by contract
 DE-AC05-06OR23177 under which the Jefferson Science Associates, LLC operate the Thomas Jefferson National Accelerator Facility, and by the grant DE-FG02-97ER41028.



\vspace{5mm}
\begin{thebibliography}{99}

\bibitem{jimayuan}
X. Ji, Jian-Ping Ma, and F. Yuan, 
{\it Phys. Rev.} {\bf D71}, 034005 (2005).

\bibitem{collinsbook}
J. C. Collins, Foundations of Perturbative QCD (Cambridge University Press, Cambridge, 2011). 


\bibitem{domarxian}	
F. Dominguez, C. Marquet, Bo-Wen Xiao, and  F. Yuan,
{\it Phys. Rev.} {\bf D83}, 105005 (2011).


\bibitem{muldrod}
P. J. Mulders and  J. Rodrigues 
{\it Phys. Rev.} {\bf D63}, 094021 (2001).




\bibitem{npb96}
I. Balitsky, 
{\it Nucl. Phys.}  {\bf B463}, 99 (1996);
{\it Phys. Rev.} {\bf D60}, 014020 (1999).

\bibitem{yura}
Yu.V. Kovchegov,  
{\it Phys. Rev.} {\bf D60}, 034008 (1999);
{\it Phys. Rev.} {\bf D61},074018 (2000).

\bibitem{muxiyu}	
A. H. Mueller, Bo-Wen Xiao, and  F. Yuan,
{\it Phys. Rev.} {\bf D88}, 114010 (2013);
{\it Phys. Rev. Lett.} {\bf 110}, 082301 (2013).

\bibitem{keld}
I. Balitsky and V.M. Braun, 
{\it Phys. Lett.} {\bf B 222}, 121 (1989);
{\it Nucl. Phys.}  {\bf B361}, 93 (1991);
{\it Nucl. Phys.}  {\bf B380}, 51 (1992).

\bibitem{bejuan}
A.V. Belitsky, X. Ji, and F. Yuan,
{\rm Nucl. Phys.}{\bf  B 656} 165 (2003).


\bibitem{mobzor}
I. Balitsky, {\it ``High-Energy QCD and Wilson Lines''}, 
In *Shifman, M. (ed.): At the frontier of particle 
physics, vol. 2*, p. 1237-1342  (World Scientific, Singapore, 2001)
[hep-ph/0101042]. 

\bibitem{nlolecture}
I. Balitsky, 
{\it ``High-Energy Ampltudes in the Next-to-Leading Order''},
in ``Subtleties in Quantum Field Theory'', ed D. Diakonov, 
(PNPI Publishing Dept., 2010)\\
arXiv:1004.0057 [hep-ph].

\bibitem{bal04}
I. Balitsky, 
{\it Phys. Rev.} {\bf D70}, 114030 (2004).
 

\bibitem{nlobk}
I. Balitsky and G.A. Chirilli,
{\it  Phys.Rev.}{\bf D77}, 014019(2008);
{\it  Nucl. Phys.}{\bf B822}, 45 (2009).


\bibitem{domumuxi}	
F. Dominguez, A.H. Mueller, S. Munier, and  Bo-Wen Xiao,
{\it Phys. Lett.} {\bf B705}, 106 (2011).

\bibitem{difope}
I. Balitsky,
{\it ``Operator expansion for diffractive high-energy scattering''},
[hep-ph/9706411].



\end{thebibliography}
\end{document}